\documentclass[pre,longbibliography,twocolumn,superscriptaddress,english,showpacs]{revtex4-1} 

\usepackage[english]{babel}
\usepackage[dvipsnames]{xcolor} 
\usepackage{graphicx} 
\usepackage{amsmath} 
\usepackage{amssymb}
\usepackage{verbatim}
\definecolor{darkblue}{rgb}{0.1,0.2,0.6} \definecolor{darkred}{rgb}{0.8,0.1,0.2}
\usepackage[colorlinks,citecolor=darkblue,linkcolor=darkred,urlcolor=darkblue]{hyperref}
\usepackage[all]{hypcap} % fix figure link problem. Links jump to top of figure now, not caption

\newcommand{\Z}{{\mathbb Z}}
\newcommand{\R}{{\mathbb R}}
\newcommand{\dd}{{\mathrm d}}

\newcommand{\ed}{{\mathrm e}}

\begin{document}

\title{Response to a small external force and fluctuations of a passive particle in a one-dimensional diffusive environment} 
\author{Fran\c{c}ois Huveneers}
\affiliation{Universit\'e Paris-Dauphine, PSL Research University, CNRS, CEREMADE, 75016 Paris, France}
\email{huveneers@ceremade.dauphine.fr}

\date{\today}

\begin{abstract} 
\noindent
We investigate the long time behavior of a passive particle evolving in a one-dimensional diffusive random environment, with diffusion constant $D$. 
We consider two cases: (a) The particle is pulled forward by a small external constant force, and (b) there is no systematic bias. 
Theoretical arguments and numerical simulations provide evidence that the particle is eventually \emph{trapped} by the environment. % both for small $D$ (quasi-static regime) and large $D$ (homogenized regime).
This is diagnosed in two ways: 
The asymptotic speed of the particle scales quadratically with the external force as it goes to zero, 
and the fluctuations scale diffusively in the unbiased environment, up to possible logarithmic corrections in both cases. 
Moreover, in the large $D$ limit (homogenized regime), we find an important transient region giving rise to other, finite-size scalings, and we describe the cross-over to the true asymptotic behavior.  
%These findings indicate that the system is genuinely out of equilibrium.  
%Our theoretical understanding can be thought of as a non-linear response theory relating the drift to the size of the fluctuations;
\end{abstract}
%\keywords{} 
%\pacs{}

\maketitle

\section{Introduction}\label{sec: introduction}
Extending the paradigms of statistical mechanics to the study of active matter is part of the main issues in contemporary theoretical physics \cite{chou_mallick_zia_2011,cates_tailleur_2015}. 
Random walks in static or dynamical random environments constitute a good case study to analyze numerous out of equilibrium phenomena
\cite{sinai_1983,sznitman_2003,basu_maes_2014,benichou_et_al_2014,guerin_et_al_2016}.
More specifically, a variety of interesting behaviors can be observed for particles advected by a viscous fluid; 
as it turns out, an initially uniform density of passive particles may display aging, clustering, phase separation and intermittency as time evolves
\cite{das_barma_2000,drossel_kardar_2000,drossel_kardar_2002,nagar_et_al_2005,nagar_et_al_2006,singha_barma_2017}. 

\begin{figure}[t]
    	\centering
   	\includegraphics[draft=false,height = 1.9cm,width = 8cm]{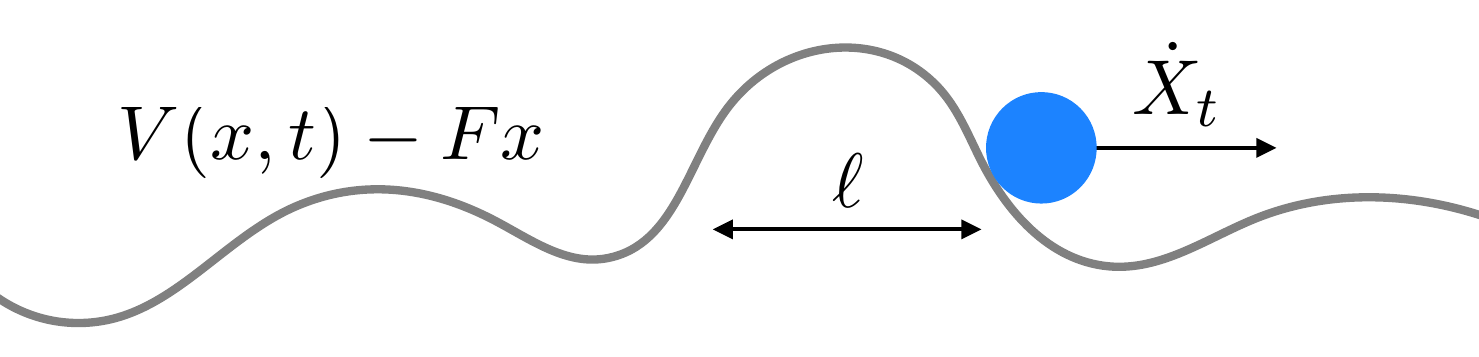}
    	\caption{Evolution of the passive particle in the potential $V(x,t) - Fx$.} 
    	\label{fig: evolution of the walker}
\end{figure}

In this paper we consider a passive particle driven by a one-dimensional ($d=1$) time dependent potential fluctuating diffusively (Edwards Wilkinson dynamics), 
as shown on Fig.~\ref{fig: evolution of the walker}. 
This system is a good candidate to host the phenomena mentioned above.  
Moreover, it is a rather natural set-up to consider since
diffusive fluctuations occur in all typical extended systems that satisfy local equilibrium and have extensive conserved quantities.
However, predicting the long time behavior of the passive particle turns out to be very puzzling in $d=1$
\cite{bohr_pikovsky_1993,gopalakrishnan_2004,avena_thomann_2012,hilario_et_al_2015,huveneers_simenhaus_2015,avena_franco_jara_vollering_2015,avena_jara_vollering_2014,avena_den_hollander_2016}
(in contrast to a lot of progress made for divergent free fields in $d\ge 2$ \cite{kraichnan_1994,falkovich_et_al_2001,fannjiang_komorowski_2002,komorowski_olla_2002}).
Indeed, 
since time correlations decay only as $t^{-d/2}$, one expects memory effects to play a dominant role in $d=1$, but it is hard to decide what their influence actually is.

Our study reveals that their role is to \emph{trap} the particle: Potential barriers confine it to a certain region of space for a finite time, and the behavior of the particle is eventually dominated by the dynamics of the barriers.
For short times, the mechanism is already visible on Fig.~\ref{fig: evolution of the walker}, while on longer timescales, it is due to the low modes of the potential; 
see \cite{sinai_1983} for the analogous phenomenology in a static environment.
In order to satisfactorily check our understanding, we consider two different set-ups and analyze them consistently.
First we analyze the differential mobility of the particle, i.e.\@ its response to a small external force, 
and second we consider its fluctuations in a unbiased environment. 
In equilibrium, these two quantities are related through the celebrated Sutherland-Einstein relation \cite{sutherland_1905,einstein_1905}, 
while generalizations of this relation to systems violating the detailed balance condition are actively studied at the present time, see
\cite{bertini_et_al_2002,komorowski_olla_2005,harada_sasa_2005,speck_seifert_2006,blickle_et_al_2007,chetrite_et_al_2008,baiesi_et_al_2009,seifert_speck_2010,baiesi_et_al_2011,lipiello_et_al_2014,sarracino_et_al_2016} 
as well as \cite{seifert_2012,ciliberto_et_al_2013}.
Our findings indicate that the system is genuinely out of equilibrium: 
The differential mobility is zero, because the asymptotic velocity of the particle scales quadratically with the applied force, while the fluctuations are normal (up to possible logarithms). 

An important aspect of the model is the presence of big finite size effects in the limit where the diffusion constant $D$ of the diffusive field grows large. 
In this regime, the trapping only becomes effective for very small external forces or very long times (depending on the considered set-up). 
This fact led to the proposal of the existence of two distinct phases as a function of $D$ in \cite{gopalakrishnan_2004}. 
We will show instead that there is a single phase and we will describe quantitatively the cross-over between a finite-size scaling region and the true asymptotic region.

The paper is organized as follows. 
After a proper description of the model in Section~\ref{sec: model}, 
we introduce the main results of this paper in Section~\ref{sec: results}, together with some brief account of previous studies. 
These results are summarized in Table~\ref{table: results}, and are shown by means of scaling arguments and numerics in 
Sections~\ref{sec: stationarity}-\ref{sec: fluctuations}. 
A heuristic theory connecting the behavior of fluctuations to the differential mobility of the particle is developed in Section \ref{sec: theory}. 
Finally Section~\ref{sec: SCA} contains several technical results on a self-consistent approximation introduced below, 
while the details of our numerical scheme are gathered in Section~\ref{sec: numerical scheme}.

\section{Model}\label{sec: model}
Let $X_t$ be the position of the particle at time $t$.  
In the overdamped regime, its evolution is governed by 
\begin{equation}\label{eq: evolution particle} 
	\dot X_t = \lambda \big(-\partial_x V (X_t ,t) + F  \big)
\end{equation}
where $\lambda$ is the mobility of the advected particle, $V$ is the fluctuating potential, and $F$ is an external constant force.
See Fig.~\ref{fig: evolution of the walker}. 
In our context, the potential $V$ may conveniently be though of as a height function and measured in units of length. 
It evolves in time according to an Edwards-Wilkinson type dynamics \cite{edwards_wilkinson_1982}
\begin{equation}\label{eq: Edwards Wilkinson}
	\partial_t V(x,t) = D \partial_x^2 V(x,t) + \xi(x,t)
\end{equation}
where $D$ is the diffusion coefficient and where $\xi$ is white noise in time and smooth in space with finite correlation length $\ell$: 
\begin{equation}\label{eq: xi xi correlator}
	\langle \xi (x,s) \xi (y,t) \rangle = 2 D \delta (t-s) \ed^{- \frac{(x-y)^2}{2 \ell^2}}.
\end{equation}
Our aim in introducing the finite correlation length $\ell$ is to avoid any problem in understanding the dynamics on short timescales, i.e.\@ $t \lesssim D^{-1}\ell^2$. 
Moreover, we assume that $X_0 = 0$ and that 
the field $-\partial_x V(x,t)$ is in equilibrium with $\langle \partial_x V(x,t) \rangle = 0$, so that $\langle X_t \rangle = 0$ at $F=0$ by symmetry. 
Numerical results are expressed in units $\ell = 1$, $D = 1$. 
To lighten the notations, we will also use these units throughout the text and mostly drop $\ell$ and $D$ from our notations
(except in a few expressions where it may just help to keep them). 
The control parameters are thus the mobility $\lambda$ and the constant force $F$.

The evolution equation \eqref{eq: evolution particle} neglects the effect of thermal fluctuations. 
Indeed, if the passive particle is at positive temperature, 
one should add some white noise term $\kappa \mathrm dB_t/\mathrm dt$ in eq.\eqref{eq: evolution particle}, where $\kappa$ is the molecular diffusivity. 
However, it is cumbersome to have an extra parameter in the model 
and one may reasonably conjecture that a finite molecular diffusivity does not affect the long time asymptotic behavior of the passive particle, 
that is eventually dictated by the low modes of the the potential.

\textbf{Fourier representation.}
Both for the theoretical analysis and numerical implementations, it is convenient to express the force field $- \partial_x V(x,t)$ by its Fourier transform: 
Let $p(\cdot)$ be the standard normal distribution (for concreteness) and 
\begin{equation}\label{eq: force field}
	-\partial_x V(x,t) = \int_\R \dd k \,  \sqrt{p(k)} \,  \big( A_k (t)\,  \ed^{i kx} + \mathrm{c.c.} \big)
\end{equation}
where $A_k(t)$ are stationary, zero mean, Gaussian processes such that  $\langle A_k (s) A_{k'} (t) \rangle =  0$ and 
\begin{equation}\label{eq: Ak correlation}
	\langle A_k (s) A_{k'}^* (t) \rangle = \frac12\delta(k-k') \ed^{-  k^2|t-s|}.
\end{equation}
%The model is thus entirely specified by the dimensionless parameters $\tilde \lambda$ and $F$. 
One can readily recover eq.(\ref{eq: Edwards Wilkinson}-\ref{eq: xi xi correlator}) from this representation, observing that
the processes $A_k(t)$ defined by the correlation \eqref{eq: Ak correlation} are independent complex Ornstein-Uhlenbeck processes obeying the evolution equation
\begin{equation}\label{eq: evolution Ak}
	\frac{\dd A_k (t)}{ \dd t} = -  k^2 A_k (t) + |k| \frac{\dd B_k (t)}{ \dd t}
\end{equation}
where $B_k(t)$ are independent complex Brownian motions: $B_k(t) = \frac{1}{\sqrt{2}}(B_k^1(t) + i B_k^2 (t))$, 
with $B_k^1(t)$ and $B_k^2 (t)$ independent real Brownian motions. 

Our numerical scheme is described in details in Section~\ref{sec: numerical scheme}, 
but it may be worth to mention here that it uses a simple discretization of the evolution equation \eqref{eq: evolution particle} 
with $- \partial_x V(x,t)$ given by eq.\eqref{eq: force field}. 
In eq.\eqref{eq: force field}, the integral is replaced by a sum over a finite number of modes. 
Since one expects the low modes to play the crucial role in determining the behavior of the particle, not all modes are sampled equally: 
the resolution becomes ever finer as $k\to 0$. 
With this way of doing, we are able to reach significantly larger times than in \cite{gopalakrishnan_2004,avena_thomann_2012}. 

\textbf{Lattice models.}
Lattice versions of the evolution equation \eqref{eq: evolution particle} have been considered both in the probabilist community \cite{avena_thomann_2012,hilario_et_al_2015,huveneers_simenhaus_2015,avena_franco_jara_vollering_2015,avena_jara_vollering_2014}
and in numerical studies, e.g.\@ \cite{gopalakrishnan_2004,nagar_et_al_2006,singha_barma_2017}.
Since our results do likely not depend on the specific modeling,
we believe that it may be of some interest to make here some explicit ``dictionary'' between our set-up and a lattice model as in \cite{huveneers_simenhaus_2015}. 

On the integer lattice $\Z$, the correlation length $\ell = 1$ featuring in eq.\eqref{eq: xi xi correlator} corresponds simply to the lattice spacing. 
As a very simple choice, one may require that the force field $- \partial_x V(x,t) = - (V(x+1,t) - V (x,t))$ takes only the values $\pm 1$, 
and that the time evolution of $V$ is governed by the so called corner flip dynamics: 
\begin{multline*}
	V(x,t+\dd t) - V(x,t) = \\ 
	\big(V(x-1,t) - 2 V(x,t) + V(x+1,t)\big) \dd N_D(t)
\end{multline*}
where $N_D(t)$ is a Poisson point process with rate $D$.  
In this case, the evolution of $-\partial_x V (x,t)$ can be mapped to the simple exclusion process, see e.g.~\cite{liggett_1985}, by identifying 
$\eta(x,t) := (-\partial_x V (x,t) + 1)/2$ with an empty site if $\eta(x,t) = 0$ and with an occupied site if  $\eta(x,t)=1$. 
The simple exclusion process is at equilibrium at a density $\rho = \langle \eta(x,t) \rangle = 0.5 + F$ for $0 \le F \le 0.5$. 
The passive particle is usually referred to as a random walker in this context, and evolves according to the following rule: 
It jumps to the right if it sits on top of an occupied site, and to the left if it sits on a vacant site, with constant jump rate $\lambda$: 
\begin{equation*}
	X_{t+\dd t} - X_t =
	 \big( \delta(\eta(X_t,t) = 1) - \delta(\eta(X_t,t) = 0) \big) \dd M_\lambda (t)
\end{equation*}
where $M_\lambda (t)$ is a Poisson point process with rate $\lambda$.

\section{Results}\label{sec: results}
The fluctuations of the passive particle for analogous or even identical set-ups have been studied in 
\cite{bohr_pikovsky_1993,gopalakrishnan_2004,nagar_et_al_2006,avena_thomann_2012}.
In \cite{bohr_pikovsky_1993}, the authors predict that $\langle X^2_t \rangle$ crosses over between a super-diffusive behavior $\lambda^2 t^{3/2}$
for $t \lesssim \lambda^{-4}$ to an almost diffusive behavior $t \log t$ for $t \gtrsim \lambda^{-4}$. 
This prediction is however not directly backed up by numerical simulations. 
In \cite{gopalakrishnan_2004}, a transition is proposed as a function of $\lambda$:
For small $\lambda$, a self-consistent approximation (SCA) predicts $\langle X_t^2 \rangle \sim (\lambda t)^{4/3}$ at long times, 
while for larger $\lambda$, the particle gets trapped by the diffusive field and becomes itself diffusive. 
We review the SCA in Section~\ref{sec: SCA}, and a close look reveals that it also predicts the behavior $\langle X_t^2 \rangle \sim \lambda^2 t^{3/2}$
for $t \lesssim \lambda^{-4}$, see eq.(\ref{variance 1 short time}-\ref{variance 1 short time particular}).
The approximations in \cite{bohr_pikovsky_1993} and \cite{gopalakrishnan_2004} for $\lambda$ small are of a similar nature, 
see eq.(\ref{eq: approx bohr pikovsky}-\ref{eq: approx gopalakrishnan}) below,
and it is not obvious to decide whether any of them is correct. 
The theoretical predictions of \cite{gopalakrishnan_2004} are validated numerically in the same paper, 
but the possibility is raised that the $(\lambda t)^{4/3}$ regime at small $\lambda$ is actually a finite size effect 
and that the particle becomes eventually always diffusive. 
In \cite{nagar_et_al_2006}, it is claimed that this is indeed what happens. 
This conclusion is however based on numerics at a single value of $\lambda$, and no hint is given that the transition does not actually occur at some lower value.  
Finally, in \cite{avena_thomann_2012}, a bunch of different exponent for $\langle X^2_t \rangle$ are observed in numerical experiments, 
but possible finite size effects are not analyzed.

Our study confirms the original prediction by \cite{bohr_pikovsky_1993}, 
though we make no clear stand on the logarithmic correction in the regime $t \gtrsim \lambda^{-4}$. 
Indeed, our data are compatible with a behavior of the type $t (\log t)^\delta$ for some $0 \le \delta < 1$, 
but we are not able to extract a value for $\delta$ and it is not even obvious whether $\delta > 0$ is a transient effect or not. 
See Fig.~\ref{fig: fluctuations}. 
In addition, we study the behavior of the walker in the presence of the external force $F>0$ in the limit $F\to 0$. 
When $F > 0$, we expect the particle to drift and eventually escape to the strong memory effects of the environment. 
For this reason, we also expect that the full system, consisting of the particle and its environment, will reach a non-equilibrium stationary states (NESS). 
We first analyze the time $T(F)$ needed for the system to reach a NESS and find that this time diverges as $F\to 0$. 
As a consequence, the system should never reach stationarity at $F=0$, which is as such a good indication that aging or trapping effects are at play. 
Second we investigate the behavior of the asymptotic velocity of the walker, 
\begin{equation}\label{eq: asymptotic velocity}
	v(F) = \lim_{t\to \infty} X_t/t \quad \text{for given} \quad F>0
\end{equation}
in the limit $F\to 0$.
Let us notice that mean field like approximations as in \cite{bohr_pikovsky_1993} or \cite{gopalakrishnan_2004} in the small $\lambda$ regime,  
would all predict the behavior $v(F) \sim \lambda F$. 
Instead we find that this scaling is only valid for $F \gtrsim \lambda$, and then crosses over to the behavior $v(F) \sim F^2$. 

All these conclusions are based on scaling arguments and numerical simulations, and are summarized in Table~\ref{table: results}. 
Two remarks are in order. 
First, the transient behavior (left column) can only be neatly observed for $\lambda$ significantly smaller than $1$, 
as a consequence of the obvious ballistic behavior of the particle for $t\le 1$. 
Second, we notice that in the true asymptotic regime (right column), the behavior of the particle does not depend anymore on the value of $\lambda$. 
This is one of the signs of the trapping by the environment.  
% \cite{leitmann_franosch_2017} seems a more different set-up even
\begin{table}[h]
	\centering
	\begin{tabular}{|c| lcl |lcl |}
		\hline
		$T(F)$ &  \multicolumn{6}{c|}{$T(F)\sim F^{-4}$} \\
		\hline
		$v(F)$ & $F > \lambda$ & : &$v(F) \sim \lambda F$ & $F <  \lambda$ &:& $v(F) \sim F^2$ \\
		\hline
		$\langle X_t^2\rangle$ & $t<\lambda^{-4}$ & : &$\langle X_t^2 \rangle \sim \lambda^2 t^{3/2}$ & $t > \lambda^{-4}$ &:& $\langle X_t^2 \rangle \sim t$\\
		\hline
	\end{tabular}
	\centering
   	\caption{Behavior of the walker, up to possible logarithmic corrections, for $\lambda \le 1$.} 
    \label{table: results}
\end{table}

Let us finally comment on a recent result in \cite{benichou_et_al_2013}: 
The variance of a tracer a particle driven by a constant force and evolving in a quasi-$1d$ diffusive environment is found to cross-over 
from a super-diffusive $t^{3/2}$ to a diffusive behavior. 
The cross-over time is there of the order of the time needed for the tracer particle to reach a new carrier (hole). 
After this time, its increments become practically independent, 
since the slow on-site decay of the correlations of the environment becomes irrelevant thanks to the drift, 
see e.g.~\cite{huveneers_simenhaus_2015} for a mathematical study of a similar phenomenology.
This results in a diffusive behavior. 
It seems thus that rater different mechanisms are at play in \cite{benichou_et_al_2013} and that the analogy with our results is mostly a coincidence.

%%\cite{leitmann_franosch}: does not seem so much related.  

\section{Time to stationarity}\label{sec: stationarity}
Let us assume that $F>0$ and let us estimate the time $T(F)$ needed for the particle to reach a stationary state, 
i.e.\@ the time after which the average of any local observable, in a frame moving with the particle, converges to some stationary value. 
As is by now well documented in the mathematical literature, in the case where the environment is itself able to relax to equilibrium in a finite time, 
$T(F)$ can be estimated, or at least upper-bounded, by this time itself, see e.g.~\cite{redig_vollering_2013} and references therein. 
This does not apply as such in our case 
since the dynamics defined by eq.\eqref{eq: Edwards Wilkinson} is diffusive and does not converge to equilibrium in a finite time, 
due to the presence of the low modes $(k \sim 0)$ in eq.\eqref{eq: force field} that relax in a time of order $1/k^2$, see eq.\eqref{eq: Ak correlation}. 
The point is however that the force $F$ provides an effective infra-red cut off for all modes with $|k| \ll F^2$. 
Indeed, the contribution of these modes corresponds roughly to the smearing of the field $- \partial_x V(x,t)$ over boxes with length of order $L \gg 1/F^2$. 
Since the amplitude of this averaged field is significantly smaller than $F$, its only effect is a slight modulation of the average velocity. 
With this cut-off, the time for stationarity of the field is of order $F^{-4}$, and we conclude that 
\begin{equation}\label{eq: time stationarity}
	T(F) \sim F^{-4} \quad \text{as} \quad F\to 0.
\end{equation}
for generic observables.

It still can be that some observables converge faster. 
Of particular interest for us is to know the time needed for the particle to reach its asymptotic speed $v(F)$ defined in eq.\eqref{eq: asymptotic velocity}. 
%(one could actually expect a time shorter than \eqref{eq: time stationarity} if the particle is not trapped by the environment).
To probe this numerically, let us measure how fast $v(F,t) := \langle X_t\rangle/t$ converges to $v(F)$ as a function of $F$, for various values of the parameter $\lambda$. 
For given $F$, let us define a rescaled time $0 \le \tau \le 1$ via $t = K F^{-4} \tau$ for some large constant $K$, and let us compare the curves
$$
	\mathcal W (\tau) = \frac{\langle X_{K F^{-4} \tau} \rangle / \tau}{\langle X_{K F^{-4} 1}\rangle/1}
$$
for various $F$. 
They collapse if the scaling \eqref{eq: time stationarity} is valid. 

\begin{figure}[t]
    	\centering
	\includegraphics[draft=false,height = 6.4cm,width = 8.4cm]{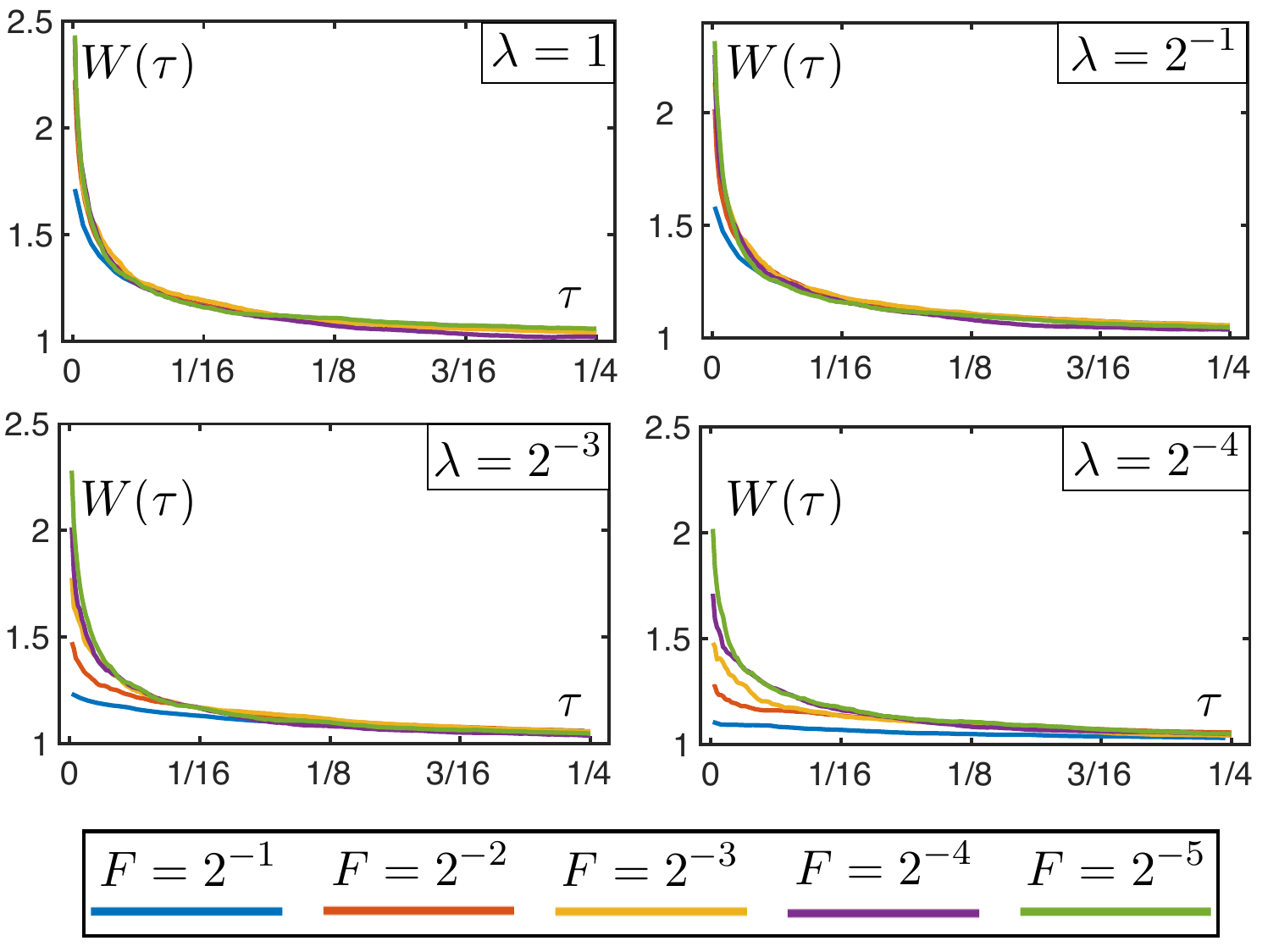}
    	\caption{	Time to stationarity:
			$W(\tau)$ for various values of $F$ and $\lambda$. 
			Average over 1000 realizations at least.} 
    	\label{fig: time to stationarity}
\end{figure}

Numerical results are shown on Fig.~\ref{fig: time to stationarity}, with $K = 2^{10}$ in the definition of the rescaled time $\tau$ 
(such a large pre-factor is needed to reach values that are stationary in good approximation at $\tau =1$). 
The scaling \eqref{eq: time stationarity} is manifestly accurate for $\lambda = 1$, $\lambda = 0.5$ (top panels).
For $\lambda = 2^{-3}$, $\lambda = 2^{-4}$ (lower panels), the scaling is only accurate for the smallest values of $F$. 
The fact that the convergence is faster than expected for the large values of 
$F$ may be interpreted as the fact that the system is initially in a state close to the stationary state.  
It is indeed reasonable that these two states are similar to each other in the homogenized regime $\lambda < F$ discussed in more details in the next section. 
The important point is thus that we observe a reasonably good collapse of the data for $F < \lambda$ in Fig.\ref{fig: time to stationarity}. 

\section{Drift}\label{sec: drift}
\begin{figure}[t]
    	\centering
	\includegraphics[draft=false,height = 10cm,width = 8cm]{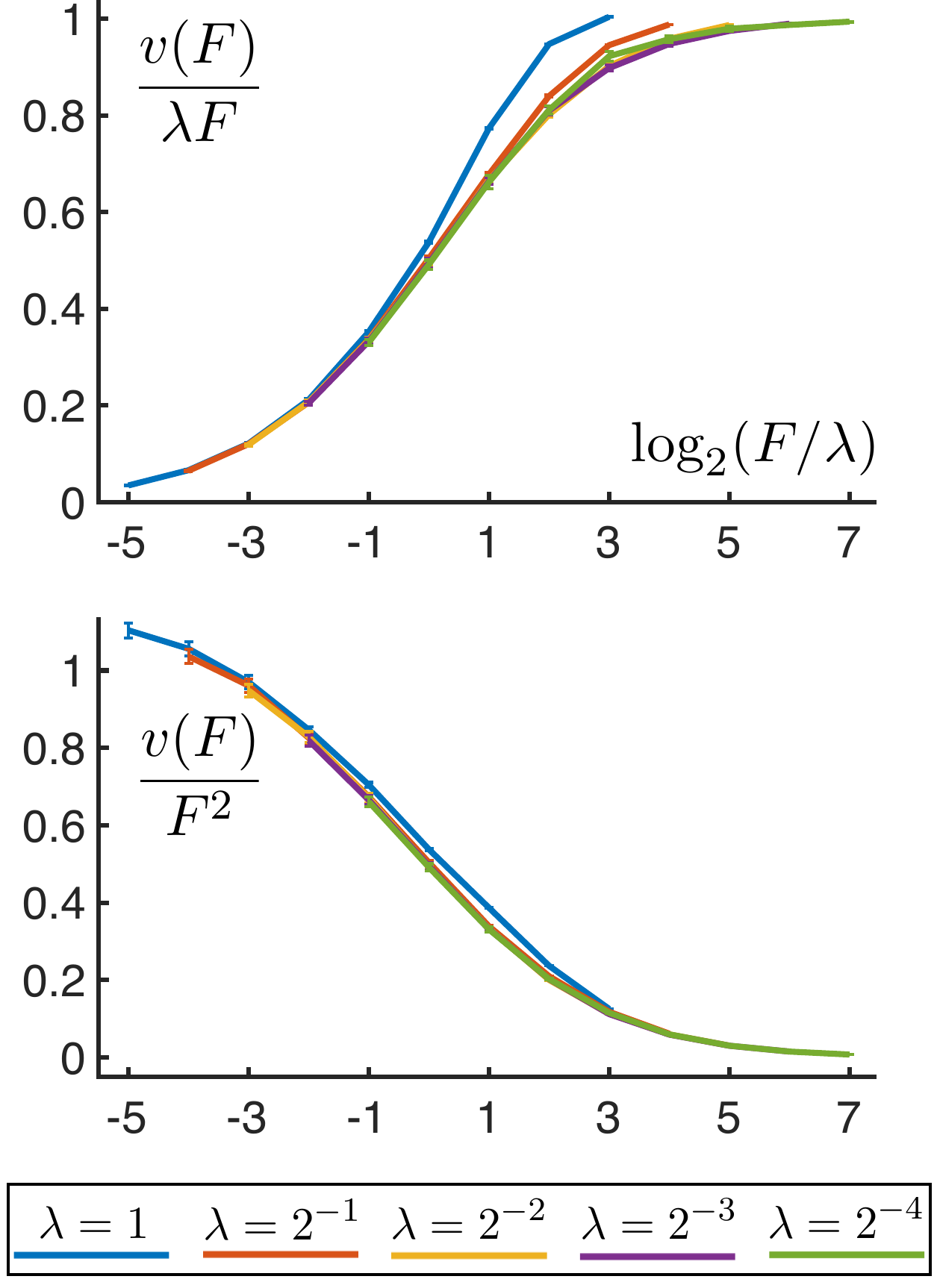}
    	\caption{	Drift:
			$v(F)/\lambda F$ (upper panel) and $v(F)/F^2$ (lower panel) as a function of $\log_2 (F/\lambda)$ for various values of $\lambda$. 
			$v(F)$ is approximated by $\langle X_{t(F)} \rangle /t(F)$ with $t(F)=2^{10}F^{-4}$ for $F = 2^{-1}, \dots , 2^{-5}$  
			(see the discussion on the time to stationarity), 
			while slightly long times are taken for $F = 2^{0}, \dots , 2^{3}$ (these points are obviously not the most relevant to determine the asymptotic $F\to 0$).
			Average over 1000 realizations at least.
	}
    	\label{fig: drift}
\end{figure}

Let us now investigate the behavior of the asymptotic velocity $v(F)$ in the limit $F \to 0$. 
The most naive expectation from eq.\eqref{eq: evolution particle} is that $v(F) \sim \lambda F$. 
This scaling is plotted on the upper panel of Fig.~\ref{fig: drift}, where one sees that it is only approximately correct for $F/\lambda > 1$.
Since our main interest is in the behavior of $v(F)$ as $F \to 0$ at a fixed value of $\lambda$, we seek thus for another scaling.
For this, we notice that the modes with $|k| \sim F^{-2}$ in eq.\eqref{eq: force field} have a strong tapping effect: 
the amplitude of this set of modes is comparable to $F$, its relaxation time is of order $F^{-4}$ and it varies in space on a scale of order $F^{-2}$.
Thus, if the field $-\partial_x V (x,t)$ would consist only of them, we would conclude right away that the velocity $v(F)$ 
of the particle must be of order $F^{-2}/F^{-4} = F^2$.
Since we identify no stronger source of slowing down, we come to the proposal $v(F) \sim F^{-2}$. 
This guess is rough and ignores the possible effects of fluctuations, stemming from all modes with momentum higher than $F^{-2}$.   
Nevertheless, 
the data in the lower panel on  Fig.~\ref{fig: drift} show that this is a reasonably good scaling, up to possible logarithmic like corrections. 

In addition, this simple way of thinking yields the cross-over value $\lambda_c \sim F$ between the two regimes represented on Fig.~\ref{fig: drift}.
Indeed, all trapping effects turn out to be prohibited for $\lambda < F$. 
The reason for this is that the modes that could trap the particle relax too fast as compared to the time needed for the particle to get trapped. 
Let us illustrate this with the set of modes with $|k| \sim F^{-2}$.
As we showed above, trapping due to these modes takes place on a length scale of order $F^{-2}$. 
But the time for the particle to travel such a distance 
(if the field $-\partial_x V(x,t)$ would consist only of these modes) 
is at least $(\lambda F)^{-1} F^{-2}$ and, in the regime $\lambda < F$, 
this is clearly larger than the relaxation time $F^{-4}$.
This reasoning can be repeated for higher modes as well (that could potentially also trap the particle though less strongly) and yields the same conclusion
(the set of modes with $|k|<F^{-2}$ has an amplitude smaller than $F$ and cannot trap the particle).

\section{Fluctuations}\label{sec: fluctuations}
Let us now set $F=0$ and study the behavior of $\langle X_t^2 \rangle$ as $t\to \infty$. 
The behavior $\langle X_t^2 \rangle \sim \lambda^2 t^{3/2}$ is best understood if one approximates
$- \partial_x V(X_t,t)$ by $-\partial_x V(0,t)$ in eq.\eqref{eq: evolution particle}, 
since this scaling follows then from an explicit computation. 
While it is clear that this approximation must be reasonable at small enough $\lambda$, it is less obvious that it is still valid up to $t \sim \lambda^{-4}$.
The kind of reasonings in \cite{bohr_pikovsky_1993,gopalakrishnan_2004} furnish eventually the shortest way to get there, 
see also eq.(\ref{variance 1 short time}-\ref{variance 1 short time particular}) below for an explicit computation. 
Moreover, in a similar way to what we did in the previous section for the drift, 
we may estimate that $t \sim \lambda^{-4}$ corresponds to the minimal time for trapping effects to appear. 
Indeed, the particle may be trapped by the modes of order $k$ if the relaxation time of these modes ($1/k^2$) 
is longer than the time needed for them to bring the particle over a distance of the order of one wavelength ($k^{-1} (k^{1/2} \lambda)^{-1}$).
Thus only modes with $k \lesssim \lambda^2$ do provide trapping, and this effect needs a time at least $\lambda^{-4}$ to be effective. 

\begin{figure}[t]
    	\centering
		\includegraphics[draft=false,height = 11.5cm,width = 8.4cm]{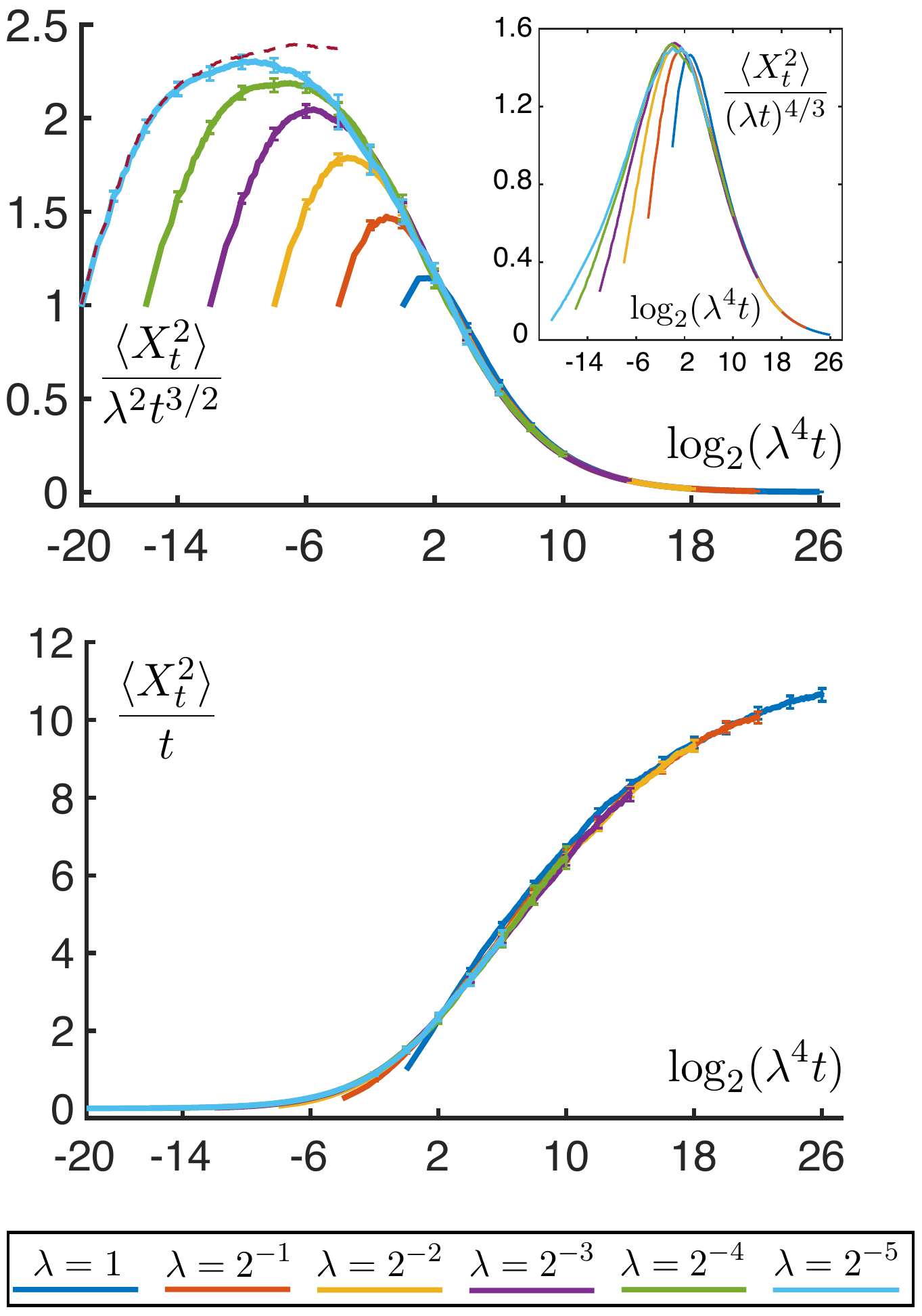}
	
    	\caption{	Fluctuations:
			$\langle X^2_t \rangle / \lambda^2 t^{3/2}$ (upper panel)
			$\langle X^2_t \rangle / t$ (lower panel)
			and $\langle X^2_t\rangle / (\lambda t)^{4/3}$ (inset)
			as a function of $\log_2 (\lambda^4 t)$ for various values of $\lambda$. 
			Dotted line in the upper panel: see the main text. 
			Average over 8000 realizations at least.} 
    	\label{fig: fluctuations}
\end{figure}

The scaling $\langle X_t^2 \rangle \sim \lambda^2 t^{3/2}$ is shown on the upper panel of Fig.~\ref{fig: fluctuations}. 
As we see, the curves do not properly collapse for $t < \lambda^{-4}$. 
The reason for this is to be found in the obvious transient ballistic behavior of the passive particle. 
To check this, we have plotted $\langle X_t^2 \rangle / \lambda^2 t^{3/2}$ for $- \partial_x V(X_t,t)$ replaced by $-\partial_x V(0,t)$ in eq.\eqref{eq: evolution particle}
at $\lambda = 2^{-5}$ 
(the result is clearly independent of $\lambda$ and this parameter only enters since the data are plotted as a function of the rescaled time $\lambda^4 t$), 
see the dotted line on the upper panel on Fig.~\ref{fig: fluctuations}.
If we could consider much smaller values of $\lambda$ and wait long enough, we would expect all curves to eventually reach a plateau at a value close to 2.5. 

The scaling $\langle X_t^2 \rangle /t$ is shown on the lower panel of Fig.~\ref{fig: fluctuations}. 
Up to logarithmic like corrections, this scaling seems accurate in the regime $t> \lambda^{-4}$. 
Finally, in the inset of Fig.~\ref{fig: fluctuations}, we consider the scaling $\langle X_t^2 \rangle / (\lambda t)^{4/3}$ predicted in \cite{gopalakrishnan_2004}. 
As we see, it is only accurate near the cross-over point $t\sim \lambda^{-4}$, where it coincides with the scalings $\lambda^2 t^{3/2}$ and $t$. 
We conclude thus that this scaling is never genuinely realized in this system. 

\section{Relating drift to fluctuations}\label{sec: theory}
We finally provide a heuristic scheme relating the behaviors of drift and fluctuations. 
This leads to the conclusion that trapping dominates the true asymptotic regime, as observed in the numerics. 
%The main thing to remember is that the origin of trapping here is eventually of a very similar nature as that of a random walker in a static environment \cite{sinai_1983}.

As a first step, we establish a phenomenological relation between the exponent $\alpha$ of the drift, and the exponent $\beta$ of the fluctuations, 
i.e.\@ 
$$
	v(F) \sim F^\alpha \; \text{as}\; F \to 0, \qquad \langle X_t^2 \rangle \sim t^\beta \; \text{as} \; t \to \infty.
$$ 
Assuming a given value for $\alpha$ (we take $1 \le \alpha \le 2$ as suggested by the data on Fig.~\ref{fig: drift}), 
we replace the evolution equation \eqref{eq: evolution particle} at $F=0$ by $\dot X_t = \lambda \varphi (X_t,t)$ where $\varphi$ is an effective force field defined by 
$$
	\varphi (x,t) = \int_\R \dd k \, | k|^{\frac{\alpha - 1}{2}} \sqrt{ p(k)} \,  \big( A_k (t)\,  \ed^{i kx} + \mathrm{c.c.} \big)
$$
with $A_k(t)$ as in \eqref{eq: Ak correlation}. % (wee keep the choice of units $\ell = 1$ and $D = 1$).
The introduction of the weight factor $| k|^{\frac{\alpha - 1}{2}}$ wrt \eqref{eq: force field} is such that 
the amplitude of the integral over $| k| \le F^2$ for any $F > 0$ is of order $F^{\alpha}$ as $F\to 0$ instead of being of order $F$ for $-\partial_x V(x,t)$ defined by \eqref{eq: force field}. 
This is consistent: If the response to an external force scales in a certain way as this force goes to zero, then the response to the lowest modes of the fluctuating field should scale the same way. 
Once the field $- \partial_x V(x,t)$ has been replaced by the effective field $\varphi (x,t)$, one may assume that all trapping effects have been taken into account and one may apply the SCA, reviewed and generalized in the next section, to determine the fluctuations of $X_t$. 
Straightforward computations yields
\begin{equation}\label{eq: alpha beta relation}
	\beta = 4 / (2 + \alpha),
\end{equation}
see eq.(\ref{variance 1}-\ref{variance 2}) below.
%generalizing our previous relations at $\alpha = 1$ and $\alpha = 2$, see SM. 
% (logarithmic corrections are actually now present at $\alpha = 2$ but we do not attribute them any reality given the level of precision of the theory), see SM. 

In a second step we determine the values of $\alpha$ and $\beta$. 
Let $T$ be some arbitrary large time, and let us decompose $\varphi$ into an almost static part and a fluctuating part, $\varphi (x,t) = \varphi_{\mathrm{sta}} (x) + \varphi_{\mathrm{flu}}(x,t) $, according to
$$
	\varphi (x,t) = \int_{|k|^2 \le 1/T} \hspace{-0.5cm}\dd k (\dots) +  \int_{|k|^2 > 1/T}  \hspace{-0.5cm}\dd k (\dots).
$$
In the absence of $\varphi_{\mathrm{flu}}(x,t)$ the particle would move to the nearest stable fixed point of $\varphi_{\mathrm{sta}}(x)$, 
i.e.\@ a point $x^*$ such that $\varphi_{\mathrm{sta}}(x^*) = 0$ and $(\dd \varphi_{\mathrm{sta}}/\dd x) (x^*) < 0 $. 
To evaluate the effect of $\varphi_{\mathrm{flu}}(x,t)$ we proceed again through the SCA; due to the infrared cut-off, fluctuations are always diffusive with a diffusion constant scaling as $T^{(2- \alpha)/4}$, see eq.\eqref{variance 3} below.
%% Precise coefficient: $D^{(2-\alpha)/4} \ell^{\alpha/2}\lambda T^{(2-\alpha)/4}$
Hence, in the vicinity of a stable fixed point $x^*$, the dynamics can be effectively described by the overdamped Ornstein-Uhlenbeck equation
$$
	\dot Y_t = - \lambda  T^{-\frac{2 + \alpha}{4}} Y_t + \lambda^{\frac{1}{2}} T^{\frac{2 - \alpha}{8}} \dd B_t/ \dd t
$$
with $Y_t = X_t - x^*$, as long as $Y_t$ remains smaller than $T^{1/2}$.  
The process $Y_t$ reaches a stationary state after some time $\tau\ll T$ for $1 \le \alpha < 2$ 
(since $\tau \sim \lambda^{-1} T^{\frac{2 + \alpha}{4}}$), 
and remarkably this state is characterized by a mean square displacement equal to $T$ for any value of $\alpha$.
Thus $X_t$ should not be trapped on a length scale $T^{1/2}$ (at least if $\alpha < 2$) but on a slightly longer length scale, 
by the same mechanism as a random walker is trapped in a static environment \cite{sinai_1983}.
Indeed $X_t$ will be trapped for a time of order $\ed^{c L^2/T}$ if $\varphi_{\mathrm{sta}}(x)$ keeps a fixed sign for length $L$.
Hence, considering an approximate mapping on the model in \cite{sinai_1983} for a lattice with spacing $T^{1/2}$ and hopping time of the walker $\tau$, 
we conclude that $X_T \sim T^{1/2} (\log T)^2$ if $\alpha < 2$ 
(the logarithmic correction is not present if $\alpha = 2$). 
% logarithmic corrections to be explained in more details at alpha = 2 actually. 
In all cases, this leads to the exponent $\beta = 1$, and hence $\alpha = 2$.

\section{Self consistent approximation}\label{sec: SCA}
We review and generalize the self-consistent approximation (SCA) introduced in \cite{gopalakrishnan_2004} yielding predictions for the average velocity and the fluctuations of the passive particle. 
Let us rewrite the evolution equation \eqref{eq: evolution particle} in integral form as
\begin{equation}\label{eq: evolution equation integral}
X_t = \lambda \int_0^t \dd s\,  \big(- \partial_x V (X_s,s) + F \big).
\end{equation}
The basic idea is to replace the process $X_t$ on the right hand side of \eqref{eq: evolution equation integral} by an independent process $Y_t$ that simply ``visits" the environment, 
in such a way that $X_t$ and $Y_t$ have the same probability distribution. 
More precisely, we look for two  processes $(X_t)_{t\ge 0}$ and $(Y_t)_{t\ge 0}$ with the three following requirements: 
(a) the processes $(X_t)_{t\ge 0}$ and $(Y_t)_{t\ge 0}$ have the same probability distribution, 
(b) the process $(Y_t)_{t\ge 0}$ is stochastically independent of the environment, hence of $(X_t)_{t\ge 0}$ (since $X_t$ depends deterministically on the environment), 
(c) $X_t$ and $Y_t$ solve the equation
\begin{equation}\label{eq: evolution equation integral Y}
X_t = \lambda \int_0^t \dd s\,  \big(- \partial V (Y_s,s) + F \big)
\end{equation}
in distribution. 
The hope is that processes $(X_t,Y_t)_{t\ge 0}$ satisfying (a-c) can be found rather explicitly 
and that the probability distribution of $X_t$ solving \eqref{eq: evolution equation integral} is qualitatively similar to the distribution of $X_t$ solving \eqref{eq: evolution equation integral Y}. 
Here, we will not deal at all with the second issue and we will solve \eqref{eq: evolution equation integral Y} at the level of the first and second moments through some Gaussian approximations (that are arguably harmless).
However, we will replace the force field $- \partial_x V (x,t)$ by some more general field $\varphi (x,t)$, as needed for the theory in Section~\ref{sec: theory}.  

\textbf{Asymptotic speed.}
For any zero average field $\varphi$ we get  $\langle X_t \rangle = \lambda F t$ by taking expectations in eq.\eqref{eq: evolution equation integral Y}. 
Hence the SCA predicts always
\begin{equation}
	v(F) = \lambda F. 
\end{equation}

\textbf{Fluctuations.}
Let us now assume $F=0$ and let us study the second moments of $X_t$. 
The generalized field $\varphi (x,t)$ is defined by eq.\eqref{eq: force field} with now
\begin{equation}
	\langle A_k (s) A_{k'}^* (t) \rangle = f(k)\delta(k-k') \ed^{- k^2|t-s|}
\end{equation}
instead of  eq.\eqref{eq: Ak correlation}, for some function $f( k)$. 
This expression boils down obviously to eq.\eqref{eq: Ak correlation} for  $f(k) = 1$,  hence  $\varphi (x,t) = - \partial_x V (x,t)$ in this case.
We will consider the more general function
\begin{equation}\label{eq: f function}
	f(k) = \chi ({|k| \ge k_0}) \,  | k|^\gamma
\end{equation}
for some $0 \le k_0 \ll 1$ and $0 \le \gamma \le 0.5$, where $\chi(A)$ is the indicator function of the set $A$.
We will look for $X_t$ and $Y_t$ having stationary increments and we will compute both $\langle X_t^2 \rangle$ in the leading order of the $t \to \infty$ asymptotic, 
and the correlations $\langle \zeta(0) \zeta (t) \rangle$, with 
\begin{equation}
	\zeta (t) = \varphi (Y_t ,t).
\end{equation} 

Let us start by computing $\langle X_t^2 \rangle$ in the large $t$ limit, without keeping track of constant prefactors:
\begin{align}
	\langle X_t^2 \rangle &=  \lambda^2 \int_0^t \int_0^t \dd s \dd s'\, \big\langle \varphi (Y_s,s) \varphi (Y_{s'},s') \big\rangle \nonumber\\
	& \sim \lambda^2 t \int_\R  \dd k \, p( k) f^2( k)    \int_{0}^t \dd \theta\, \ed^{-k^2 \theta} \langle \cos k Y_\theta \rangle \nonumber\\
	& \sim \lambda^2 t \int_0^t \dd \theta \int_\R  \dd k \, \ed^{- k^2 (\frac{1}{2} +  \theta + \langle X_\theta^2 \rangle)} f^2( k) .
	\label{eq: SC equation for variance}
\end{align}
To get the second line, we used the assumption that $(Y_t)_{t\ge 0}$ have stationary increments; 
to get the last line, we used that $p(z)$ is a standard normal distribution, that  $\langle X_t^2 \rangle = \langle Y_t^2 \rangle$ by assumption, and that $(Y_t)_{t \ge 0}$ is Gaussian.
This last assumption is presumably not exact, but we expect that the results do not depend qualitatively on this Gaussian approximation. 
Equation \eqref{eq: SC equation for variance} is the self-consistent equation solved by the variance $\langle X_t^2\rangle$. 
For various values of $k_0$ and $\gamma$ in eq.\eqref{eq: f function}, eq.\eqref{eq: SC equation for variance} yields
\begin{align} 
	&\langle X_t^2 \rangle \sim (\lambda t)^{\frac{4}{3+2\gamma}},  \qquad 0 \le \gamma < 0.5, \; k_0 = 0, \label{variance 1}\\
	&\langle X_t^2 \rangle \sim \lambda t (\log t)^{\frac{1}{2}},  \qquad \gamma = 0.5, \; k_0 = 0, \label{variance 2}\\
	&\langle X_t^2 \rangle \sim k_0^{\gamma - \frac12} \lambda t, \qquad 0 \le \gamma < 0.5, \; k_0 >0.\label{variance 3}
\end{align}

Next, to compute the correlations $\langle \zeta(0) \zeta(t)  \rangle$, we notice that 
\begin{align}
	\langle \zeta(0) \zeta(t) \rangle 
	&= \langle \varphi(Y(0),0) \varphi(Y(t),t) \rangle \nonumber\\
	&\sim \int\dd k\, f^2 (k) \ed^{-k^2 (\frac{1}{2} + t + \langle X_t^2 \rangle)}.
\end{align}
Hence, from (\ref{variance 1}-\ref{variance 3}), 
\begin{align}
	&\langle \zeta(0) \zeta (t)  \rangle  \sim    \frac1{ (\lambda t)^{2 \frac{1+2\gamma}{3+2\gamma}}} , \quad 0 \le \gamma < 0.5,\; k_0 = 0,\label{correlation 1}\\
	&\langle \zeta(0) \zeta (t)  \rangle  \sim \frac1{(\lambda t) (\log t)^{\frac{1}{2}}} , \quad \gamma = 0.5,\; k_0 = 0, \label{correlation 2}\\
	&\langle \zeta(0) \zeta (t)  \rangle \sim   \frac{\ed^{- k_0^{\frac32 + \gamma}(\lambda t)}}{(\ell k_0)^{\gamma^2 - \frac14} (\lambda t)^{\frac12 + \gamma}}, \nonumber\\ 
	&\hspace*{4cm} 0 \le \gamma < 0.5,\; k_0 >0. \label{correlation 3}
\end{align}

\textbf{Remark on transient behavior.}
The expressions (\ref{variance 1}-\ref{variance 3}) and (\ref{correlation 1}-\ref{correlation 3}) 
only hold in the limit $t \to \infty$ at fixed values of all other parameters, and may have to be modified on some transient time scales. 
E.g.\@ to determine eq.\eqref{variance 1}, we assumed that $ \theta \lesssim \langle X_\theta^2 \rangle$ in eq.\eqref{eq: SC equation for variance}. 
If this condition is violated, we obtain instead 
\begin{equation}\label{variance 1 short time}
	\langle X_t^2 \rangle \sim \lambda^2 t^{2 - \frac{1+\gamma}{2}}
\end{equation}
and in particular $\langle X_t^2 \rangle \sim \lambda^2 t^{3/2}$ for $\gamma = 0$.  
This expression should replace eq.\eqref{variance 1} for all times short enough so that $t \gtrsim \langle X_t^2 \rangle$ 
for $ \langle X_t^2 \rangle$ given by eq.\eqref{variance 1 short time}. 
E.g.\@ for $\gamma = 0$ we find that eq.\eqref{variance 1 short time} holds as long as
\begin{equation}\label{variance 1 short time particular}
t \lesssim \lambda^{-4}.  
\end{equation}
We recover thus the behavior announced in Section~\ref{sec: results}.

\textbf{Remark on static environments.}
The same SCA can be applied to a random walker in a static environment ($D = 0$) in $d=1$:
\begin{equation}\label{eq: static case}
	\dot X_t = -\lambda \partial_x V(X_t) + \kappa \frac{\dd B_t}{ \dd t } 
\end{equation}
(one needs here to take the molecular diffusivity $\kappa$ to be finite in order to avoid a trivial dynamics). 
If we reintroduce the parameters $D,\ell$ in eq.\eqref{variance 1}, we find actually $\langle X_t^2 \rangle \sim \ell^2 (\lambda t /\ell)^{4/3}$.
This expression is independent of $D$, and performing similar computations for the evolution equation \eqref{eq: static case} yields again the same expression. 
In this case, it is known that the SCA does not predict the correct behavior for $\langle X_t^2 \rangle$ at any value of $\lambda$. 
Indeed the particle is always strongly sub-diffusive: $\langle X_t ^2\rangle$ scales as $ (\log t)^4$ as $t\to \infty$, see \cite{sinai_1983}.

\textbf{Remark on \cite{bohr_pikovsky_1993} and \cite{gopalakrishnan_2004}.}
It may be interesting to point out explicitly the difference between the approximations made in \cite{bohr_pikovsky_1993} and \cite{gopalakrishnan_2004}
to determine the asymptotic behavior of $\langle X_t^2 \rangle$ as $t\to \infty$ for $F=0$.
In both cases, it is determined through a consistency condition, but the correlator 
$C_{s,s'} = \langle \partial_x V (X_s,s) \partial_x V (X_{s'},s')\rangle$ 
is estimated in a slightly different way. Let us assume $s' \ge s$. 
In \cite{bohr_pikovsky_1993}, the approximation
\begin{equation}\label{eq: approx bohr pikovsky}
	C_{s,s'} \sim \langle \partial_x V (\langle X^2_{s'-s}\rangle^{1/2},s'-s) \partial_x V (0,0)\rangle
\end{equation}
is used, while the slightly more refined approximation 
\begin{equation}\label{eq: approx gopalakrishnan}
	C_{s,s'} \sim \int \dd y \frac{\ed^{-y^2/ 2\langle X_{s'-s}^2 \rangle}}{\sqrt{2 \pi \langle X_{s'-s}^2\rangle}}
	\langle \partial_x V(y,s'-s) \partial_x V(0,0) \rangle. 
\end{equation}
is made in \cite{gopalakrishnan_2004}, for $s' - s > 0$.
%This little twist has striking consequences.  

\section{Numerical scheme}\label{sec: numerical scheme} 
We describe the discretization of eq.\eqref{eq: evolution particle} and (\ref{eq: force field}-\ref{eq: Ak correlation}) used in our numerics. 
Let us denote by $\Delta t$ the elementary time step of the particle. 
Eq.\eqref{eq: evolution particle} becomes
\begin{equation}\label{eq: discretized evolution}
	X_{t+\Delta t} = X_t + \lambda (\Delta t) (- \partial_x V(X_t,t) + F).
\end{equation}
The integral \eqref{eq: force field} defining $- \partial_x V (x,t)$ becomes a sum 
\begin{equation}
	- \partial_x V (x,t) = \sum_{k \in K} \rho(k) (A_k(t) \ed^{i kx} + \mathrm{c.c.})
\end{equation}
where $K = \{k_1, \dots , k_N\}$ is the set of accessible inverse wavelength $k_i > 0$ for $1 \le i \le N$, and where $\rho(k)$ is the weight of mode $k$. 
Given some $0 < \delta < 1$, we set
\begin{equation}
	k_i = \delta^{i - 1}, \quad 1 \le i \le N
\end{equation}
and the corresponding weights
\begin{equation}
	\rho (k_i) = (\delta^{i-1} - \delta^i)^{\frac{1}{2}}, \quad 1 \le i \le N- 1
\end{equation}
and $\rho(k_N) = \delta^{(N-1)/2}$. 

Let us next see how the force field $-\partial_x V(x,t)$ is updated. 
Since every mode $A_k(t)$ is an independent Ornstein-Uhlenbeck process evolving according to eq.\eqref{eq: evolution Ak},
if one knows the value of $A_k(t)$ at some time $t$, 
one can write explicitly the value of the real and imaginary part of $A_k(t')$ at any time $t' > t$: 
\begin{multline}\label{eq: update A k t prime}
	\Re A_k (t') = \ed^{-k^2 (t' - t)} \Re A_k (t) \\
	+ \mathcal N \left(0, \frac{1 - \ed^{-2 k^2 (t'-t)}}{4}\right)
\end{multline}
and similarly for the imaginary part (real and imaginary part are independent), 
where $\mathcal N (m, \sigma^2)$ denotes a normal distribution with mean $m$ and variance $\sigma^2$. 
In our scheme, there is no reason to update the modes at times shorter than the elementary time step $\Delta t$ of the particle. 
Moreover, it is reasonable to update the low modes less frequently than the high modes, in such a way that all modes are updated in essentially the same way each time they are. 
The mode $k_i$ is updated once every   
\begin{equation}
	\Delta t_i = \left\lceil K k_i^{-2} \right\rceil \, \Delta t
\end{equation}
for some $K > 0$, as
\begin{multline}
	\Re A_k (t + \Delta t_i) = \ed^{-D k_i^2 \Delta t_i} \Re A_k (t)  \\
	+ \mathcal N\left(0, \frac{1 - \ed^{-2D k_i^2 \Delta t_i}}{4}\right), 
\end{multline}
and similarly for the imaginary part. 
As we see, if not for rounding off, the update is identical for all modes since $D k_i^2 \Delta t_i \simeq D K \Delta t$. 

%The very large and very small $\lambda$ regimes are not accessed in the same way: 
%to reach small values of $\lambda$ one may fix $D$ and decrease $\lambda$ to ever smaller values, or one may alternatively fix $\lambda$ and increase $D$; 
%instead to reach large values of $\lambda$, it would be wrong to keep increasing $\lambda$, and the only way is to decrease $D$. 
%In this work we focussed on small and moderate values of $\lambda$, and we decided to fix $D$ and vary $\lambda$ between $0$ and $1$.

Since $\Delta t$ comes as $(\Delta t) \lambda$ in eq.\eqref{eq: discretized evolution}, one may fix $\Delta t = 1$. 
%Moreover, by changing units, one may fix $\ell = 1$ and $D = 1$. 
%By changing units, one may fix $\Delta t = 1$ and $\ell = 1$. Moreover, since the problem depends on $\lambda$ and $D$ through $\tilde\lambda$, one may fix $D = 1$. 
The parameters $N,\delta$ and $K$ are discretization parameters, fixed to $N = 130$, $\delta = 0.9$ and $K = - \log (0.5) \simeq 0.7$ in all our experiments.
With these values of $N$ and $\delta$, our results should be safe of any periodicity or quasi-periodicity effects. 
Moreover, for intermediate time scales, we checked that reasonable variations of these parameters did not fundamentally affect the results.

\section{Conclusions}\label{sec: conclusion}
We have investigated the behavior of a passive particle advected by a fluctuating surface in the Edwards-Wilkinson universality class. 
Both the differential mobility and the fluctuations have been analyzed with the same rational.  
Our study exhibits the existence of a finite size scaling limit, that differs from the true asymptotic limit.
The latter regime is dominated by trapping effects of the environment. 
%We hope that our work may clarify the status of some previous findings \cite{gopalakrishnan_2004,avena_thomann_2012}.
%In the future, we would like to extend our methodology to other environments with different roughness exponents, such as the KPZ surfaces considered in \cite{drossel_kardar_2000,drossel_kardar_2002}. 

\begin{acknowledgments} 
	I thank M.~Barma, M.~Salvi, F.~Simenhaus, T.~Singha, G.~Stoltz and F.~V\"ollering for helpful and stimulating discussions. 
	I also thank an anonymous referee for very useful suggestions. 
	I benefited from the support of the projects EDNHS ANR-14-CE25-0011 and LSD ANR-15-CE40-0020-01 of the French National Research Agency (ANR). 
\end{acknowledgments}

\bibliography{RW_bibliography}

\end{document}